# Designing Fair AI for Managing Employees in Organizations: A Review, Critique, and Design Agenda

**Lionel P. Robert[1], Casey Pierce[1], Liz Morris[1], Sangmi Kim[1], Rasha Alahmad[1]**

*[1]University of Michigan*

lprobert@umich.edu; cbspierc@umich.edu; ebmorris@umich.edu; sangmik@umich.edu; rashama@umich.edu

**Running Head: Designing Fair AI for Managing Employees**

## Abstract

Organizations are rapidly deploying artificial intelligence (AI) systems to manage their workers. However, AI has been found at times to be unfair to workers. Unfairness toward workers has been associated with decreased worker effort and increased worker turnover. To avoid such problems, AI systems must be designed to support fairness and redress instances of unfairness. Despite the attention related to AI unfairness, there has not been a theoretical and systematic approach to developing a design agenda. This paper addresses the issue in three ways. First, we introduce the organizational justice theory, three different fairness types (distributive, procedural, interactional), and the frameworks for redressing instances of unfairness (retributive justice, restorative justice). Second, we review the design literature that specifically focuses on issues of AI fairness in organizations. Third, we propose a design agenda for AI fairness in organizations that applies each of the fairness types to organizational scenarios. Then, the paper concludes with implications for future research.

# 1. INTRODUCTION

Organizations are rapidly employing artificial intelligence (AI) to manage their workers (Gerlsbeck, 2018; Hughes et al., 2019; Kolbjørnsrud, Amico, & Thomas, 2016). AI systems— computer systems that can sense, reason, and respond to their environment in real time, often with human-like intelligence—have the potential to help organizations efficiently and effectively manage their employees. Organizations are turning to AI to direct, supervise, and hold their workers accountable (Hughes et al., 2019; Jarrahi, 2018; Rosenblat, 2018). For example, in a survey of 1,770 managers from 14 counties, 86% of executives stated that they planned to use AI for managing their employees, including monitoring, coordinating, and controlling their workers (Kolbjørnsrud et al., 2016). Furthermore, the survey reported that 78% of the managers trust decisions made by AI (Kolbjørnsrud et al., 2016). As these reports indicate, there is increased interest in using AI for managing work in organizations.

However, decisions and actions by an AI system might not be fair to the employees (Harini, 2018). This is in large part because of the biases encoded directly into the AI or the ways in which the AI learns from human behavior over time. AI systems can make millions of decisions independently of a human decision-maker; however, those decisions can be unpredictable and invisible to the worker who is being impacted by those decisions (Martin, 2019). Echoing these concerns, John Giannandrea, who leads AI at Google, warned that AI shaped by human prejudices is far more dangerous than potential killer-robots (Knight, 2017). For example, Amazon recently discovered its AI-powered recruitment engine exhibited bias against female applicants when compared to male applicants (Gonzalez, 2018). As such, workers might experience unfairness in organizations because of AI, which could in turn result in decreased worker effort and increased worker turnover (Greenberg & Colquitt, 2005). Therefore, if AI is to effectively manage workers, it must be designed to support fairness and even redress instances of unfairness.

Despite the attention directed toward AI unfairness, a lack of literature is focused on developing a theoretical and systematic design agenda for fair AI. This is problematic for several reasons. First, the current discussion on the topic often fails to acknowledge the distinct types of fairness, such as the fairness of outcomes, process, or interactions. Ignoring the differences among types of fairness can lead to a one-size-fits-all approach when designing for AI fairness in organizations. Second, the lack of an overarching theoretical framework makes it difficult to guide, organize, and integrate design solutions across the broader HCI community. This results in a fragmented and incoherent view of both the problem and design space related to AI fairness. Nonetheless, AI must be designed to be fair and just if it is to effectively manage workers.

To address issues of AI fairness in organizations, in this paper we contribute to the HCI community in the following ways. First, we present a theoretical framework that leverages the organizational justice theory. The theoretical framework both provides conceptual clarity on the various types of unfairness and helps to organize, integrate, and guide current conversations on AI fairness. Second, we present a review of the scant design literature on AI fairness in organizations. We organize the current literature according to our theoretical framework. In doing so, we highlight shortcomings of the literature on AI fairness that inform design. Finally, we introduce a framework to guide our design research agenda for fairer AI in organizations.

## 2. THEORETICAL BACKGROUND

### 2.1. Artificial Intelligence (AI) in Organizations

AI is used in organizations to plan, organize, control, and lead. The movement of AI from research laboratories to organizations is not new. Early applications of AI in organizations were in the form of expert and knowledge management systems (Doukidis & Paul, 1990). These applications were designed primarily to aid in human decision-making (Duchessi, O'Keefe, & O'Leary, 1993). As such, these applications had little autonomy or agency of their own and were employed for a narrow set of tasks. However, the ability to capture, store, and rapidly process large data sets in real time has expanded the use and the degree of autonomy associated with AI in organizations. For example, AI is now being used to automatically search, filter, select, and recommend job applicants (Daugherty, Wilson, & Chowdhury, 2019). AI systems are not only automatically assigning workers to tasks but are also evaluating their work (Dhir & Chhabra, 2019; Hoshino, Slobodin, & Bernoudy, 2018). These work activities occur daily with minimal human intervention or awareness (Rosenblat, 2018).

The expanded use and increased autonomy of AI can be problematic. Previously, organizations underestimated the extent to which human biases could impact AI systems. For example, an early paper highlighted the advantage of AI to organizations by arguing that AI offers "more consistent decision making and greater reliability in decision processes" (Duchessi et al., 1993, p. 154). Although misguided, this rhetoric has not been entirely discarded. For example, as noted in a recent *Harvard Business Review* article, AI can be "potentially more impartial in its actions than human beings" (Kolbjørnsrud et al., 2016, p. 5). However, scholars generally now acknowledge that AI is far from less susceptible to human prejudices or biases (Daugherty et al., 2019).

Examining fairness, or the lack thereof, in AI is becoming recognized as an increasingly important topic of investigation across many research communities. The ACM Conference on Fairness, Accountability, and Transparency (FAT), which started in 2018, is one example of this recognition. Scholars from diverse fields such as management, law, information, data science, medical science, and public policy are questioning and examining issues of AI fairness (Jarrahi, 2018; Siau & Wang, 2018; Veale & Brass, 2019; Vellido, 2019; Žliobaitė, 2017). However, without a theoretical and systematic approach, researchers and designers lack a road map for crafting interventions likely to help promote fairer AI. In the next section, we introduce a theoretical framework to help address these issues.

## 2.2.   From Equity Theory to Organizational Justice Theory

Adams' (1963, 1965) equity theory is by far the most widely used theory to explain fairness in the organizational literature (Virtanen & Elovainio, 2018). At its core, Adams' equity theory proposes that the distribution of rewards should be based on an individual's contribution (Roch et al., 2019). Adams' equity theory is largely based on the concept of reciprocity—workers contribute effort and the organization reciprocates those efforts. If workers contribute more, they receive more rewards; if workers contribute the same, they receive the same rewards (Ye, You, & Robert, 2017). Taken together, Adams' equity theory speaks directly to the social contract that exists between workers and their employers.

Organizational justice theory leverages and extends Adams' equity theory. Specifically, organizational justice theory is concerned with explaining workers' reactions to unfairness or inequities (Greenberg, 1990; Virtanen & Elovainio, 2018). Whereas Adams' equity theory helps explain when workers might believe a situation is unfair, organizational justice theory explains how workers might react to unfairness (Greenberg & Colquitt, 2005; Pérez-Rodríguez, Topa, &

Beléndez, 2019). For example, the literature on organizational justice theory has shown that when workers believe they are being treated unfairly they often attempt to resolve it by contributing less to their organization or even by leaving their organization (Cowherd & Levine, 1992; Hopkins & Weathington, 2006). The literature on organizational justice theory has identified three widely used types of fairness: distributive, procedural, and interactional (Greenberg & Colquitt, 2005). Table 1 provides a description of each fairness type.

**Table 1**
**Fairness and Redressing Unfairness Types**

| Name | Definition |
| --- | --- |
| **Types of Fairness** | |
| Distributive | Fairness with respect to the allocation of outcomes such as pay and other resources |
| Procedural | Fairness defined by the process employed to reach or decide an outcome |
| Interactional | Fairness defined by how employees are treated by their organization |
|    - Interpersonal | Degree that employees are treated with respect and dignity |
|    - Informational | Degree that information is provided to help employees understand processes taken to achieve fairness and their outcomes |
| **Types of Justice (Redressing Unfairness)** | |
| Restorative | Focuses on making the offended party or victim whole again |
| Retributive | Focuses on punishing the offender in relation to the severity of the offense |

## 2.3.   Defining Fairness Types

**Distributive fairness**

Distributive fairness refers to fairness with respect to the allocation of outcomes such as pay and other resources (Alexander & Ruderman, 1987). Distributive fairness aligns closely with Adams' equity theory (Cook & Hegtvedt, 1983; Greenberg, 1990). Employees assess distributive fairness by comparing their work contributions against their pay or reward from work (Greenberg, 1990). For example, distributive fairness is achieved when 8 hours of work is compensated by workers receiving pay for 8 hours rather than 7 hours of work. The former would be viewed as matching the outcome with workers' input while the latter would not. Distributive fairness can be assessed by not only comparing a worker's work to his or her payment but the payment to that of other workers performing the same work (Greenberg, 1990). Employees can believe the work situation is unfair when others doing the same work receive more payment. Distributive fairness is often at the heart of gender discrimination suits (e.g., Chan, 2000), where women are paid significantly less than men for the same work.

**Procedural fairness**

Procedural fairness is defined by the process employed to reach or decide the final outcome. Procedural fairness is achieved through processes (e.g., resolving disputes, distributing resources) that are visible to the employee (i.e., transparent) and that are consistent both across people and over time (He, Zhu, & Zheng, 2014; Levanthal, 1980). To be procedurally fair, all parties must be heard before a decision is made (Leventhal, 1980). Six characteristics have been used to describe whether procedures are fair: consistency, unbiased suppression, representativeness, correctability, accuracy, and ethicality (Leventhal, Karuza, & Fry, 1980). For example, a grievance process that allows workers to be heard by a third party to dispute an employment termination would be seen as procedurally fairer than a process that only allows workers to appeal to the supervisor who terminated them. The former would be viewed as potentially less biased than the latter.

**Interactional fairness**

Interactional fairness is defined by how workers are treated by their organization (Luo, 2007).

Interactional fairness is achieved by organizations ensuring that workers are both well respected and well informed (Baron, 1993; Cohen-Charash & Spector, 2001). Interactional fairness has two aspects: interpersonal fairness and informational fairness (Colquitt, Conlon, Wesson, Porter,

& Ng, 2001). Interpersonal fairness refers to the degree of treating others with respect and dignity (Cropanzano, Bowen, & Gilliland, 2007). Informational fairness refers to the explanations that are provided to individuals to help them understand what processes were undertaken to achieve fairness and what the outcomes of those processes were. Researchers (e.g.,

Brockner & Wiesenfeld, 1996) have argued that interactional fairness occurs when workers are

treated with dignity (Baron, 1993) and workers are provided with explanations and clarifications

(Bobocel & Farrell, 1996; Shapiro, Buttner, & Barry, 1994). For example, a manager displays

interactional fairness when sharing information appropriately and respectfully (Cropanzano et al., 2007). Many reports of injustice reference the ways in which workers were treated interpersonally during interactions (Mikula, Petri, & Tanzer, 1990).

   Although all three types of fairness have been linked to similar positive work outcomes

(e.g., organizational trust and commitment, job performance, and satisfaction), they are

conceptually distinct from one another (Colquitt et al., 2001; Cropanzano et al., 2007; McFarlin

& Sweeney, 1992; Poon, 2012; Rupp & Cropanzano, 2002; van Dijke, De Cremer, Mayer, &

Van Quaquebeke, 2012). For example, a female worker might believe she is receiving unfair payment for her work (i.e., distributive fairness). The organization may or may not provide a fair process for her to appeal and request a pay increase (i.e., procedural fairness). Throughout the entire ordeal her manager might begin to treat her rudely and disrespectfully in response to her appeal (i.e., interactional fairness). Each of these illustrates a different work arrangement and lens through which to evaluate fairness (Cropanzano, 2001).

## 2.4. Redressing Unfairness

Even in organizations that prioritize fairness, workers might still experience instances of unfairness. Despite this organizational reality, little attention has focused on achieving justice[1]—that is, the approaches to redress unfairness (Darley & Pittman, 2003). Specifically, two approaches to redress unfairness—restorative justice and retributive justice—provide ways to deal with the consequences of unfairness in organizations (Bradfield & Aquino, 1999; Darley & Pittman, 2003; Wenzel, Okimoto, Feather, & Platow, 2008).

**Restorative justice**

Restorative justice focuses on making the offended party or victim whole again. First introduced into the organizational literature by Bradfield and Aquino (1999), the goal of restorative justice is to heal damaged relationships—often through making amends, apologies, and forgiveness—and reintegrate offenders back into the organization where the offender and victim work (Goodstein & Aquino, 2010). Rooted in indigenous cultures' dispute resolution methods, restorative justice gained momentum in the criminology literature as a way to address the shortcomings of due process in Western-style legal systems (Kidder, 2007).

Practicing restorative justice requires three main values: participation, reparation, and reintegration (Kidder, 2007). All affected parties must participate in a meeting or meetings to resolve the conflict. In organizations, three stakeholders are involved in restorative justice: the offender, the victim, and the wider organizational community (Goodstein & Aquino, 2010; Sabag & Schmitt, 2016). During this meeting, both the victim and the offender have the opportunity to express their side of the story and how they were affected by the offense. Through this meeting, reparations are decided upon to fix the damage the offense caused. Finally, the offender is reintegrated back into the wider community of the organization—often through forgiveness by the victim (Kidder, 2007).

---

[1] While the literature uses the terms *fairness* and *justice* interchangeably, in this paper we use the term *fairness* in reference to the three fairness types (distributive, procedural, interactional) and the term *justice* in reference to the ways in which actors redress unfairness (restorative, retributive).

**Retributive justice**

Retributive justice focuses on punishing the offender in relation to the severity of the offense (Sabag & Schmitt, 2016; Wenzel et al., 2008). Unlike restorative justice, there is little participation by the offender or the victim; instead unilateral punishment is given to an offender, often by a neutral third party (Wenzel et al., 2008). Retributive justice does not require compensation to the victim (Darley & Pittman, 2003). An example of this type of justice is the Western-style legal system, where the offense is evaluated by a jury and punished by a judge (Sabag & Schmitt, 2016).

Organizations might punish the offender or withdraw a desired reward from a worker who has committed an offense (Trevino, 1992). Generally, offenses perceived as humiliating or disempowering make retributive justice salient (Wenzel et al., 2008). This is particularly true when the offense is perceived to have been intentional, which produces moral outrage and a desire to see the offender punished (Darley & Pittman, 2003). However, much of the organizational literature discourages punishing subordinates because it is assumed to have negative outcomes that outweigh the benefits (Luthans & Kreitner, 1985). Aspects of the punishment, such as harshness, have been shown to negatively impact attitudinal outcomes such as trust in supervisors and organizational commitment, and future evaluations of procedural and distributive justice (Ball, Trevino, & Sims, 1993). Nonetheless, when workers perceive that punishment will follow an action, they are less likely to commit that action (Willison, Warkentin, & Johnston, 2018).

The simplest way to clearly distinguish restorative and retributive justice from the other types of justice is to view them as what happens after unfairness has been established. For instance, in the previous example of a female worker's wage appeal, the organizational policy that allows for an appeal refers to the processes to determine whether she was unfairly paid (i.e., procedural fairness). However, once it is determined that she was unfairly paid relative to other comparable workers (i.e., distributive fairness), the next step would be to redress the unfairness (i.e., restorative justice or retributive justice). One approach would be to simply increase the female worker's salary and give her back pay of her lost wages (i.e., restorative justice). Another approach, which is not mutually exclusive, would be to punish those responsible for the payment disparity (i.e., retributive justice). Table 1 provides a description of both types of justice for redressing unfairness.

## 3. Literature Review on Artificial Intelligence (AI) Fairness in Organizations

We conducted a literature review to examine how organizational justice theory is used in the AI fairness literature. As a part of this process, we focused our efforts on AI fairness papers that gave design implications and that directly related to an organizational work context. These choices served our goal of both reviewing the current state of the design-focused AI fairness literature and developing a design agenda for fair AI in organizations based on organizational justice theory. We found 61 articles on AI fairness, using a keyword search on several search engines, and screened the articles based on our inclusion and exclusion criteria. To be included, the study had to discuss one of the fairness types, AI fairness in the context of organizations or workplaces, and design implications of AI fairness. As a result, a total of 25 papers published between 2008 and 2019 were included in our literature review. Our processes of selecting, screening, and coding AI fairness studies are described in detail in the appendix.

## 3.1. Overview of Literature on Designing AI for Organizations

In this section, we review how the AI fairness design literature has used distributive, procedural, and interactional fairness. In our review of the AI fairness design literature, we found a lack of coverage of retributive justice and restorative justice. Thus, the following section does not include a discussion on retributive and restorative justices. We also summarize the papers that did not use any of the aforementioned fairness definitions but still informed our understanding of AI fairness. Finally, we discuss the shortcomings of the current literature as it relates to organizational justice theory.

**AI and distributive fairness**

Distributive fairness was the most commonly discussed type of fairness in our literature review, included in 18 of 25 (72%) total design papers (see Table 2). The emphasis on distributive fairness might be expected because many of these papers were focused on designing fair decision-making systems that allocate outcomes such as organizational resources (Lee & Baykal, 2017) and who gets selected for a position (Mouzannar, Ohannessian, & Srebro, 2019). Although the term *distributive fairness* was only explicitly used in five papers (Brown, Chouldechova, Putnam-Hornstein, Tobin, & Vaithianathan, 2019; Grgić-Hlača, Zafar, Gummadi, & Weller, 2018; Lee et al., 2018; Martin, 2018; Ötting & Maier, 2018), we were able to categorize 14 other papers that employed an approach to fairness that followed the traditional definition of distributive fairness (i.e., fair allocation of outcomes).

Although distributive fairness was the most commonly discussed type of fairness, there was variation in the depth in these papers regarding organizational justice theory, some drawing on it heavily (e.g., Brown et al., 2019; Grgić-Hlača et al., 2018; Ötting & Maier, 2018) and others mentioning it more briefly in a larger discussion of fairness (e.g., Lee & Baykal, 2017; Martin, 2018). Overall, the AI literature only examined one of the traditional ways of assessing distributive fairness, and within this traditional assessment approach, fairness was largely implemented in one of two ways: through mathematically validated algorithms and through standards set by legal documents.

**Table 2**
**Summary of Included Literature Review Papers, 2008–2019**

| Organizational Justice | | Time Period | | | |
| --- | --- | --- | --- | --- | --- |
| | | **2008** | **2017** | **2018** | **2019** |
| **Types of Fairness** | **Distributive** | 1 | 1 | 7 | 9 |
| | **Procedural** | | | 6 | 6 |
| | **Interactional** | 1 | | 3 | 2 |
| **Types of Justice (Redressing Unfairness)** | **Restorative/Retributive** | | | | |
| **Number of Papers** | | 2 | 1 | 16 | 17 |

In the organizational justice literature, distributive justice is often assessed either by comparing an individual's inputs to their outputs (i.e., within individuals) or by comparing an individual's outcomes to others' outcomes (i.e., between individuals or groups). In our review of the AI fairness literature, we found that distributive fairness was overwhelmingly measured by comparing outcomes across people rather within the individual (e.g., Glymour & Herington, 2018; Grgić-Hlača et al., 2018; Heidari & Krause, 2018). For example, Heidari and Krause (2018) argued that fairness should be measured by determining whether similar individuals should be assigned similar outcomes by the algorithm. This focus on assessing fairness among individuals or groups might be from the AI fairness literature's tendency to measure distributed fairness from the outside (i.e., detecting disparate impact of outcomes via a mathematically validated algorithm or legal standard) rather than from the perspective of the impacted individual.

To date, the primary method of achieving fair outcomes across individuals has been through choosing a computational model that operationalizes equity with a goal of fairly distributing the outcome for which the AI is responsible (e.g., De Jong, Tuyls, & Verbeeck, 2008; Madras, Creager, Pitassi, & Zemel, 2019; Mouzannar et al., 2019; Zhang & Bareinboim, 2018). These were either based on mathematically validated definitions (De Jong et al., 2008; Glymour & Herington, 2019) or legal principles that prohibit discriminatory distributions of outcomes (Heidari & Krause, 2018; Kyriazanos, Thanos, & Thomopoulos, 2019; Mouzannar et al., 2019). Non-empirical papers (Abrams, Abrams, Cullen, & Goldstein, 2019; Hacker, 2018) proposed further recommendations for integrating existing discrimination and data protection laws into AI systems. For example, Hacker (2018, p.1146) proposed the combination of law with algorithms as facilitating "equal treatment by design."

**AI and procedural fairness**

Procedural fairness was the next commonly discussed type of fairness, described in 12 of 25 (48%) total design papers (see Table 2). These papers primarily focused on employing AI to automate existing procedures rather than employing AI to create new procedures. These papers suggested that designers uphold two of the main components of procedural fairness—consistency and transparency—to support an organization's existing processes for fair decision-making (Brown et al., 2019; Grgić-Hlača et al., 2018; Hu, 2018; Ötting & Maier, 2018; Zhang &

Bareinboim, 2018). In this way, AI applications should be predictable in that they follow the same procedure every time and be transparent in how this procedure is enacted through the algorithm.

One main theme in the procedural fairness articles we reviewed was automating the decision-making in existing procedures (Abrams et al., 2019; Lee, 2018; Ötting & Maier, 2018). Authors argued that to have a procedurally fair AI, the organization's policies and processes about decision-making need to be developed to be procedurally fair before they are automated (Abrams et al., 2019; Brown et al., 2019). This argument is supported by empirical evidence that procedural justice is important regardless of the type of agent (human or AI) that makes a given decision (Ötting & Maier, 2018). Additionally, the type of procedure is important because users have reported perceiving procedural fairness to be high when an AI system makes an objective decision and low when it makes a more human, subjective decision (Lee, 2018). This is unsurprising because the difficulty of designing to account for complex social processes has been noted (Hu, 2018; Selbst, Boyd, Friedler, Venkatasubramanian, & Vertesi, 2019). The procedure being automated should be carefully chosen and assessed for fairness because otherwise the AI is simply reifying procedures that have already put some people at a disadvantage.

After a fair and acceptable procedure has been chosen to be automated through AI, authors emphasized the importance of the AI performing that particular procedure in a consistent and transparent way (Brown et al., 2019; Oswald, 2018). Transparency and consistency are important because people who are impacted by the decisions made by AI have reported needing to know how the algorithms work before they can trust the decisions they make (Brown et al., 2019). One suggestion of how to achieve transparency and consistency could be through applying existing policy about transparency to AI (Oswald, 2018). However, few papers explicitly described how to go about taking an existing procedure and translating it into AI transparently and consistently.

**AI and interactional fairness**

Interactional fairness was the least common type of fairness in our literature review, included in 6 of 25 (24%) total papers (see Table 2). These papers could be divided into those that primarily focused on informational fairness, those that primarily focused on interpersonal fairness, and those that focused on both. Terms like transparency, along with comprehensive and unbiased presentation of information, were used in papers we categorized as informational fairness (Abrams et al., 2019; Janzen, Bleymehl, Alam, Xu, & Stein, 2018; Ötting & Maier, 2018). For example, Abrams et al. (2019) presented and discussed an ethical framework along with values and core principles to guide the design and use of AI. They highlighted the importance of fair interactions between people and AI and defined fair interactions as the presentation of complete and accurate information.

Terms like cooperation and respect were used in the papers we categorized as interpersonal fairness (Brown et al., 2019; Daugherty et al., 2019; De Jong et al., 2008; Janzen et al., 2018). For example, Janzen et al. (2018) focused on interpersonal fairness between a human and a dialogue system. They referred to a type of equitable satisfaction based on both cooperative and competitive motives—cooperative in that the dialogue system wants to provide the customer with comprehensive information regarding a product, and competitive in that the customer wants to buy the best product at the best price, which might result in a loss of sale for the dialogue system.

**Table 3**
**Summary of Literature Review by Fairness Type**

| | Citation | Fairness Type | | | Justice Type (Redressing Unfairness) |
|---|---|---|---|---|---|
| | | Distributive | Procedural | Interactional | Restorative /Retributive |
| 1 | Abrams et al. (2019) | X | X | X | |
| 2 | Brown et al. (2019) | X | X | X | |
| 3 | Daugherty et al. (2019) | | | X | |
| 4 | De Jong et al. (2008) | X | | X | |
| 5 | Glymour & Herington (2019) | X | X | | |
| 6 | Grgić-Hlača et al. (2018) | X | X | | |
| 7 | Hacker (2018) | X | X | | |
| 8 | Heidari & Krause (2018) | X | | | |
| 9 | Holstein et al. (2018) | | | | |
| 10 | Holstein et al. (2019) | | | | |
| 11 | Hu (2018) | X | X | | |
| 12 | Hutchinson & Mitchell (2018) | | | | |
| 13 | Janzen et al. (2018) | X | | X | |
| 14 | Kyriazanos et al. (2019) | X | | | |
| 15 | Lee & Baykal (2017) | X | | | |
| 16 | Lee (2018) | | X | | |
| 17 | Madras et al. (2019) | X | | | |
| 18 | Martin (2018) | X | X | | |
| 19 | Mouzannar et al. (2019) | X | | | |
| 20 | Oswald (2018) | | X | | |
| 21 | Ötting & Maier (2018) | X | X | X | |
| 22 | Rastegarpanah et al. (2019) | | | | |
| 23 | Selbst et al. (2019) | X | X | | |
| 24 | Veale et al. (2018) | X | | | |
| 25 | Zhang & Bareinboim (2018) | X | X | | |
| | **Number of Papers** | **18 (72%)** | **12 (48%)** | **6 (24%)** | **0 (0%)** |

Daugherty et al. (2019) also discussed how AI can be unfair in its interactions with minorities. Daugherty et al. used examples of how AI responded to or failed to respond to individuals based on their skin color or the types of voices an AI application employed during its interpersonal interactions with people. Please see Table 3 for a list of all papers by fairness type.

**Other literature: model and data selection**

Our literature review also revealed ways of assessing fairness that did not easily align with our organizational justice framework. Specifically, these included algorithmic-based activities that focused on the design or training of the algorithm as the method of achieving fairness. Few of these papers in our "other literature" category mentioned the goal of promoting any of the organizational justice fairness types, such as distributive fairness (e.g., Grgić-Hlača et al., 2018); instead, these papers focused on how fairness was a consequence of either the structure (i.e., selection of the underlying computational model for the algorithm) or the training of the algorithm itself (i.e., data selection).

One way of assessing fairness that emerged from our literature review surrounds the computational model of the algorithm. From this perspective, fairness is achieved through encoding a chosen quantitative definition, or computational model of fairness into the algorithm. These computational models are usually based on existing statistical or behavioral economics theories (e.g., De Jong et al., 2008; Glymour & Herington, 2019; Madras et al., 2019), derivations based on non-mathematical sources (e.g., Grgić-Hlača et al., 2018; Hacker, 2018; Kyriazanos et al., 2019), or developing frameworks that monitor fairness over time (e.g., Heidari & Krause, 2018; Mouzannar et al., 2019).

The other method of assessing fairness that emerged from our literature review surrounds the data used to train the algorithm. This perspective argues that it is imperative to select the appropriate dataset to train the algorithm to avoid unwanted biases in the AI system (Daugherty et al., 2019). However, obtaining the appropriate dataset for the given organizational context is not a trivial or simple process. First, designers must assess whether the datasets that they are considering using to train the algorithm have any unwanted biases reflected in that data (Abrams et al., 2019; Rastegarpanah, Gummadi, & Crovella, 2019). Next, designers might still need to correct for unwanted bias by including additional data in the dataset (Rastegarpanah et al., 2019), weighting the dataset, or curating new datasets that more accurately represent the organizational context in which the AI will be deployed (Daugherty et al., 2019). Taking these steps has been shown to improve the fairness of recommender systems by reducing the variation and outliers that often drive unfairness and improving the accuracy of the models (Rastegarpanah et al., 2019).

## 3.2.    Shortcomings of the Existing Literature

Although few articles in our literature review explicitly discussed distributive, procedural, and interactional fairness, we were able to categorize most of the articles into our theoretical framework. However, the existing AI fairness literature presents some shortcomings. Before we discuss them it should be noted that the attention directed toward AI fairness is increasing, as shown in Figure 1. That being said, these shortcomings are mainly a lack of differentiation among types of fairness, a lack of consideration for the organizational context that surrounds AI systems and impacts fairness, and little to no coverage of how to redress instances of unfairness after they occur.

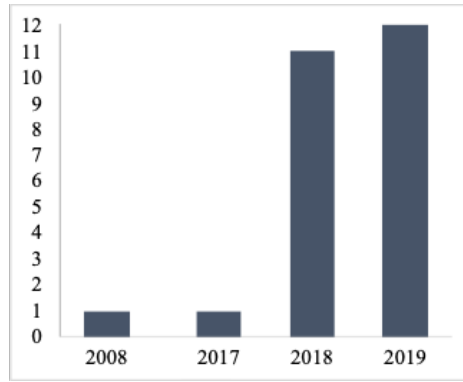

**Figure 1. Number of AI fairness publications in literature review of 2008–2019**

When discussing AI fairness, the articles we reviewed rarely differentiated among types of fairness and were often vague in their definitions of fairness. For example, Abrams and colleagues (2019) defined fair as when "data are used in a manner most useful to stakeholders while mitigating risks as much as possible" (p. 22). This lack of specificity and differentiation led to an overreliance on measuring fairness as it relates to outcomes (i.e., distributive fairness). As stated in the previous section on AI and distributed fairness, this could be because of the ability to assess fairness by comparing outcomes across individuals with relative ease. This might also be a result of the lack of awareness that there are other types of fairness. For example, distributive fairness aligns well with most common and popular definitions of fairness. In either case, this does not negate the importance and need to address procedural and interactional fairness. This is particularly true for interactional fairness, which was underrepresented in the literature that we reviewed. This underrepresentation is particularly problematic because a large portion of injustices reported by workers are related to their interpersonal interactions (Mikula et al., 1990). A lack of consideration for interactional fairness can result in poor experiences and negative worker outcomes.

The lack of focus on interactional fairness might be contributing to the lack of consideration for the organizational or social context surrounding the use of the AI system.

However, there were some notable exceptions that considered stakeholders' perceptions of the AI system (e.g., Brown et al., 2019). As Cropanzano (2001) noted, the organizational context impacts which types of fairness are most important at any given time. For example, a platform company, where employees are managed exclusively through AI (e.g., an app on their phone), is a very different organizational context than a more traditional company where there are both human and AI interactions. In the platform company context, paying attention to interactional fairness might be particularly important, because the AI is the employee's primary interaction with the company and other employees. Procedural fairness would also be important to make sure the employee knows how the work should be done. In the traditional organizational context, however, distributive fairness might be of particular interest because workers could compare their wages to one another. Thus, it is important to consider organizational context in order to attend to the appropriate fairness type for any given situation to successfully manage employees.

Finally, the articles we reviewed did not consider how the AI system should handle instances of unfairness. This oversight in the literature fails to address the organizational realities that even under the best circumstances and intentions, unfairness can occur. However, merely re-establishing distributive, procedural, and interactional fairness is not sufficient to restore justice to those who have been treated unfairly (Wenzel et al., 2008). Explicitly redressing instances of unfairness in the design of AI systems can create organizational norms that support seeking justice for employees (Kidder, 2007) and also help organizations regain legitimacy when unfairness has occurred (Pfarrer, Decelles, Smith, & Taylor, 2008).

## 4.  Design Agenda for Artificial Intelligence (AI) Fairness in Organizations

In the following sections we present a design agenda that applies each of the fairness types to AI systems in organizations. The goal of this framework is not to prescribe a particular fairness type that organizations should pursue, nor to claim that one fairness type is more suitable than another. Instead, we posit that fairness should not be approached arbitrarily as a broad term, but rather designers should consider the specific tenets of each fairness type when creating AI systems for organizations. We acknowledge that AI systems cannot resolve all existing or future unfairness in organizations. Nonetheless, we introduce our design agenda to identify potential design issues and stimulate discussion that is imperative to consider for designing fair AI systems.

Next we propose a framework to support AI fairness that includes four primary components and four affordances (see Table 4 and Table 5). The four primary components include: (1) operationalizing fairness, (2) selecting appropriate datasets and computational models, (3) evaluating fairness, and (4) redressing instances of unfairness. The four affordances that support AI fairness are transparency, explainability, voice, and visualization. Drawing from Leonardi and Treem's (2012) articulation of affordances, we define affordances as the particular ways in which an individual perceives and interacts with a system. By considering affordances, this paper acknowledges that employees are users of AI who come to understand how AI both constrains or enables possibilities for actions. In doing so, this paper hopes to avoid a deterministic view of AI that seems to presuppose that users are only being acted upon and do not interact with the technology. Additionally, we also adopt an affordance approach in our framework of AI fairness, rather than providing an exhaustive list of features, because specific features become obsolete as technologies change or in different organizational contexts. Nonetheless, we do provide examples of features to illustrate how employees can enact these AI affordances.

## 4.1. Framework to Support AI Fairness in Organizations

**Operationalizing fairness**

The selecting, defining, and specifying of fairness and how it will be evaluated can be used to define operationalizing fairness. Considering that technologies are not value-neutral, it is imperative that organizations operationalize their management practices before delegating those practices to AI. Operationalizing fairness first requires organizations to identify the specific management practice (e.g., hiring, promotion, compensation) and then determine how they will define that practice according to a specific fairness type. Given the nuances in how each fairness type is defined, operationalization also pushes organizations to move beyond vague statements of fairness and consider how to enact fairness in practice.

**Table 4**
**Proposed Framework to Support Fairness**

| Primary Components | | Definitions |
|---|---|---|
| 1. Operationalizing Fairness | | The selecting, defining, and specifying of how fairness will be evaluated |
| 2. Data and Model Fairness | Data Measurement Bias | Bias caused by how something was actually measured |
| | Data Variable Collection Bias | Bias caused by the decision to include or exclude specific variables in the data set |
| | Data Sample Bias | Bias caused by the data sample not being representative of the broader population that will interact with the AI |
| | Model Bias | Bias occurring when inappropriate variables are selected and appropriate variables are not, or when the variables selected are assigned inappropriate weights |
| 3. Evaluating Fairness | | The determination of AI fairness |
| 4. Redressing Unfairness | | Actions taken to remedy or set right unfairness |

*Note.* AI = artificial intelligence

**Data and model fairness**

As previously discussed, our literature review uncovered papers that did not directly address fairness through the lens of the organizational justice literature, but mathematically represented fairness by applying data and computational models. We include this perspective in our proposed framework to discuss how data and model biases can directly impact the ways in which each of the fairness types are enacted through AI systems in organizations.

*Data measurement bias*

Defined as bias caused by how something was actually measured, data measurement bias is one source of potential bias. Data measurement is often the result of human biases. For example, employees' performance evaluation score is likely to be the result of not only their work performance but their relationship with their rating manager. In addition, their performance evaluation score is also likely to be influenced by other contextual factors such as co-worker support and organizational support.

*Data variable collection bias*

This bias is caused by the decision to include or exclude specific variables in the data set. Because the model seeks to maximize prediction based on the variables included in the dataset, the decision to include or exclude specific variables can have profound impacts on the variables eventually selected in the model. Therefore, bias can occur through decisions made prior to any analysis. In many cases, the variables selected are not based on sound or thoughtful decision-making but rather on convenience or availability. Organizations rely on the data in hand rather than holding off their analysis until new appropriate and relevant variables can be collected.

*Data sample bias*

Bias caused by the data sample not being representative of the broader population that the AI will interact with is defined as data sample bias. The data sample can be biased by either over- or under-representing certain groups. When the data sample is not representative, the model produced from it is likely to assign more weight to factors that are more predictive of the over-represented group. As such, models based on such data are also likely to be better at predicting the over-represented groups and much poorer at predicting the under-represented groups. In addition, the time frame used to determine which data should be included or excluded can be a source of data sample bias.

*Model bias*

Model bias occurs when inappropriate variables are selected and appropriate variables are not, or the variables selected are assigned inappropriate weights. Models are developed to achieve predictive power and not to follow a particular logical reasoning. Empirically these models are developed based on variable correlations rather than causation. This means models can leverage any variable that helps to increase their predictive power. This can lead models to include variables that are normally considered illegal or unacceptable (race, age, gender). Even with interventions to exclude such variables from the model, other variables that are highly correlated with them can still be included into the model and act as proxy variables.

**Evaluating fairness outcomes**

Evaluating fairness outcomes involves the actual determination of whether the AI was fair. Using the fairness type as a guiding framework to evaluate the fairness of the system, organizations need to continually evaluate whether the system did in fact produce fair outcomes. For example, if an organization used an AI system to determine employee compensation, the organization should evaluate whether employees who work the same number of hours in the same position are paid equally regardless of gender, race, or age. If the organization finds significant differences across demographics, this could be evidence suggesting that the AI system is producing unfair compensation outcomes for employees and that the AI system needs to be updated accordingly.

**Redressing instances of unfairness**

Finally, even though achieving fairness might be the organization's goal, AI system users can still experience unfairness. Redressing unfairness speaks directly to the actions taken to remedy or set right unfairness. In our review of the literature, we found that redressing instances of unfairness through AI systems has been understudied. Thus, we propose that when considering designing for algorithmic fairness, designers need to consider how to redress instances of unfairness before implementing AI systems in organizations. Specifically, AI systems should be able to determine whether unfairness has occurred and also afford employees a path to redress an instance of unfairness through restorative justice or retributive justice.

## 4.2. Affordances to Support Fairness in AI Systems

One of the common critiques about AI systems is the opaque nature of the AI "black box" that makes it difficult for employees to fully understand how their data are being used or to interact with the technology as informed users. To these ends, to guide design decisions for AI fairness we propose four affordances: transparency, explainability, visualization, and voice. We also discuss various ways to evaluate the effectiveness of these affordances.

Table 5

*Artificial Intelligence Affordances*

| Affordance | Definition | Examples of Features | Effectiveness Measurement |
|---|---|---|---|
| Transparency | Making the underlying AI mechanics visible and known to the employee | Protected, private log-in feature that provides employees access their data<br><br>Feature that notifies the employee when data are being collected or have changed | Interpretability |
| Explainability | Describing the AI's decision/actions to the employee in human terms | Voice or text features that allow the AI to converse with employees about a decision | Interpretability |
| Visualization | Representing information to employees via images, diagrams, or animations | Features that allow the AI to represent data such as charts and figures<br><br>Animation features that enable employees to simulate possible "if/then" scenarios based on variables of interest | Interpretability |
| Voice | Providing employees with an opportunity to communicate and provide feedback to the AI | Features that allow the employee to flag biased or inaccurate data<br><br>Features that allow employees to comment on their agreement or disagreement with the AI's decisions and actions<br><br>Features that allow employees to rate their comprehension or agree with the AI's decisions and actions | Input influence |

*Note.* AI = artificial intelligence

## Transparency

The transparency affordance focuses on making the underlying AI mechanics evident—that is, being both visible and known—to the employee. Transparency affords the employee the ability to see what data are being used (data selection), how those data are being computed (model selection) in the AI system, and for what management purposes the AI system is being used (e.g., promotion decisions, compensation evaluations, performance reviews). For example, a feature providing employees with a private and secure view of their data that are used in the AI system could provide employees visibility to the decision criteria used in their evaluation decisions. Additionally, notification features could make it known to employees when changes have been made to those decision criteria. Such feature examples illustrate how AI can support transparency in making the AI mechanisms concerning data selection both visible and known to employees.

**Explainability**

The explainability affordance refers to the capability of the AI to describe its decisions and actions to the employee in human terms. This means that the AI is able to explain how its decisions and actions are related to management activities (e.g., evaluating performance, calculating compensation). In this way, explainability is important for employees to understand how a particular management activity is being enacted through AI. Furthermore, explainability is important with regard to the AI describing to employees when and how their behaviors at work are being monitored. For example, avatar or chatbot features could use text or voice to converse with employees. These features, which could explain to employees a recent decision on receiving a new compensation or being declined for a promotion, for example, might be useful to inform employees about how the AI determined such decisions.

**Visualization**

Visualization affordance is the capability of the AI to represent information to employees via images, diagrams, or animations. One of the reasons organizations are increasingly delegating management practices to AI systems is to leverage large amounts of data. However, it is not always clear how to make sense of these data for individual employees. Thus, the visualization affordance considers the ways in which AI system interfaces can communicate data clearly and accurately to employees. For example, this can include features such as affording employees options to select variables of interest concerning their employment outcomes (e.g., compensation, performance evaluations) to generate a visual representation of their employment data over time. Another visualization feature might afford employees the ability to simulate future outcomes based on their data. For instance, an employee might want to see a graph that shows how the likelihood for a promotion or compensation increase could change based on factors such as years of experience, education, or opting to take maternity/paternity leave.

**Voice**

The voice affordance provides employees with an opportunity to communicate with and give feedback to the AI. In this way, employees can interact with AI systems by having a chance to indicate when they believe the AI has treated them unfairly or when they do not understand the AI mechanisms, which is necessary to redress instances of AI unfairness. For instance, a flagging feature could provide employees the ability to indicate when they find inaccurate data about their work performance or outcomes. Additionally, commenting or rating features could provide employees opportunities to express how they believe AI decisions have been biased or inaccurate. Such documented feedback through these feedback features could help designers to better understand employees' experiences and interactions with AI, thus in turn understanding how end-users interpret how AI processes and decisions might be unfair. The affordances are outlined in Table 5.

## 4.3. Measuring effectiveness of affordances

The aforementioned affordances should also be measured for effectiveness to ensure that they are indeed supporting AI fairness. Specifically, transparency, explainability, and visualization affordances should be measured in terms of interpretability—that is, the extent to which employees are able to understand the AI's decisions and processes. The voice affordance should be measured in terms of input influence, or the extent to which employees feel empowered to contribute to the AI.

**Interpretability**

First, with regard to the transparency, explainability, and visualization affordances, it is important to consider the realities that all employees might not have the expertise to understand the information presented to them as end-users of the AI. Being able to interpret the data inputs and outputs of the AI system provides employees with the agency to act upon that information. Thus, interpretability as an effectiveness measure serves as a check for designers to ensure that employees understand what the AI is doing and how AI mechanisms impact employees' experiences and outcomes. Otherwise, when employees are unable to interpret AI, they lack an understanding of whether and how the AI has treated them unfairly; if that is the case, when employees are treated unfairly it is difficult for them to seek a path toward restorative or retributive justice. As such, interpretability is an important measure of transparency effectiveness for users to understand whether they are being treated fairly given the outcomes of the AI.

**Input influence**

Second, with regard to the voice affordance, features that enable employees to communicate when they do not understand the AI or when they perceive that the AI has treated them unfairly are only valuable to employees if the employees believe they are actually being heard. Thus, input influence as an effectiveness measure of the voice affordance is determined not only by the employees' ability to contribute to the AI, but also the extent to which employees believe their feedback and suggestions to the AI can influence the AI actions and decisions. Otherwise, employees might feel as if their contributions are meaningless and that the AI is acting upon them with no recourse for action if unfairness occurs. Therefore, input influence as a measure of effectiveness of the voice affordance allows designers to consider whether employees feel empowered to contribute and believe that their contributions to AI are actually being heard.

## 4.4.  Applying AI Fairness Framework to Fairness Types

Using the framework outlined in the prior sections, we introduce three fictional problem cases to illustrate how to apply the framework for each fairness type. Our goal is not to present a definitive conclusion on how to resolve each case, but rather to highlight the challenges and questions that designers need to consider when developing AI for organizations. Next we present an example problem case for each fairness type.

**AI fairness example case: Distributive fairness**

Distributive fairness is often discussed with regard to determining employee compensation. Imagine that Bob and Sue are equally qualified for a pay raise in their roles as department managers at a sales company. They both have worked the same length of time and generally have had the same level of education, skill and performance. However, Sue has taken 6 months for maternity leave in the last year. Although Sue might have improved her department's sales record before taking her maternity leave, if her performance record is compared to Bob's over the last year, she would be disadvantaged for a pay raise. Recognizing that employees like Sue who take maternity or paternity leave might be treated unfairly in this regard, the company could correct for distributive unfairness by instituting a policy requiring human resources not to count maternity or paternity leave time against an employees' sales performance record.

If the company were to use an AI system to automate how they allocate pay raises throughout the company, this might provide advantages for the company to predict high performers or systemize who receives pay increases without relying on promotion committees. Nonetheless, AI systems could perpetuate potential issues with regard to gender discrimination in pay decisions. This could be in part due to the nature of the AI system that makes use of existing biased data to calculate and recommend decisions. As such, it would also be difficult to detect whether the discrepancies in how pay increases are distributed through the company are caused by the complexity and lack of clarity of the AI estimation.

*Operationalizing fairness*

The first step in operationalizing distributive fairness involves deciding the particular fairness outcome. For instance, as in the case presented here, worker compensation could focus on different outcomes, such as health care coverage, vacation days, opportunities for promotion, or paid personal time. For the purposes of our problem case, we focused on equal pay as the fairness outcome. As such, we then operationalized fair pay based on (1) whether the pay matches the work effort contributed and (2) how an employee's pay compares to the pay other employees have received.

*Data and model fairness*

Operationalizing pay outcomes using distributive fairness raises two issues in selecting the appropriate data and models in an AI system: (1) What data should count as worker efforts? and (2) What standards should be used when comparing workers' efforts to their outcomes? In the case of Sue's pay promotion, the current dataset used to make pay increase decisions might not have an explicit gender designation for employees. However, other criteria used to make compensation decisions, such as overall sales, attendance to sales meetings, and additional training might be correlated with gender such as women being more likely than men to take a maternity/paternity leave. For example, because Sue took 6 months of maternity leave, she showed fewer annual sales, attended fewer sales meetings and was able to complete fewer training sessions in the last year than Bob, who worked 12 months. Thus, one potential option when selecting appropriate data for pay decisions would be to make sure the AI system accounts for time-sensitive data, such that the time an employee takes for maternity or paternity leave would not be accounted for as unproductive time so that it would not disadvantage employees like Sue when they were being considered for a pay increase. Unless biased data and employee history were corrected in the dataset, the future decisions on pay increases for Sue would reflect biased productivity data and perpetuate unfair treatment.

*Evaluating distributive fairness*

Reflecting on the approach in how we would operationalize distributive fairness in this case, we would first evaluate whether Sue's pay increase (or the lack thereof) matched the work effort she contributed to her sales and how her pay outcomes compared to her colleagues'. This raises two important issues in evaluating distributive fairness in AI systems: (1) What comparison groups should be used? and (2) What information is needed to ensure that the comparison groups are indeed valid?

As illustrated in our problem case, comparing the rate of women selected for a pay increase with rate of men selected for a pay increase is one method to evaluate whether distributive fairness was achieved. This approach often materializes in the form of comparing outcomes among members of social categories and assumes that a good comparison group is available. Discovering that women are much less likely to be selected for promotion than men can trigger a review of the selection criteria. This approach can also leverage other types of comparisons of groups not based on social categories. An example of this is found in the literature by Saxena et al. (2018), who referred to this approach as "treating similar individuals similarly" (p. 2). By doing so, it also assumes that all other differences between groups are accounted for and that group members represent most of the differences in outcomes. In Sue's case, this might mean that the AI system uses other employees who have taken a maternity or paternity leave for a similar length of time as a more appropriate comparison group to determine whether distributive fairness was achieved.

*Redressing instances of distributive unfairness*

In theory, the simplest approach for an AI system to redress distributive unfairness is to restore back to the employee the outcome that he or she should have received. As suggested in the case, if the AI system determined that Sue did not receive a pay promotion that she would have received if she had not taken maternity leave, the AI system could predict what her back pay would have been and recommend that Human Resources adjust her pay accordingly. Or the AI system could prioritize Sue being eligible for a pay raise when returning from maternity leave or a promotion commensurable with her past performance that would pay her at the fair rate.

However, restorative outcomes in practice might be difficult to determine. For example, Sue might have been unfairly withheld a pay promotion, thus losing out on future earnings but also enduring psychological and emotional suffering. Hence, in this particular case, it might be difficult for an AI system to determine the monetary value of mental and emotional damages to fully restore the worker to justice. This in turn raises several important issues: (1) How can an AI system restore justice for outcomes that are difficult to quantify? and (2) How does an AI system determine the value of those hard-to-quantify outcomes? These issues raise important considerations regarding whether AI systems should quantify and automate such organizational decisions. For one, the organization should determine whether it will quantify difficult outcomes like mental and emotional distress because this could raise future issues and complications assessing distributive fairness. As our problem case illustrates, there might be multiple options for how the AI system could restore justice to Sue, and it would be difficult for organizations to determine how to automate those decisions in an AI system. Although there is not always a simple solution for restoring justice in this regard, the significant point to recognize is that organizations should include some viable options for restoring justice rather than completely ignoring any opportunities to redress instances of distributive unfairness.

*Affordances to support distributive fairness*

Transparency, explainability, and visualization are the three primary affordances that designers should consider when creating AI systems to support distributive fairness. Because distributive fairness involves comparing one person's individual outcomes relative to others' outcomes, having access to data is important to measure comparative outcomes. As such, designers should address specific questions related to: (1) the level of data the AI provides to make meaningful comparisons to assess whether someone has been treated fairly and (2) how much data should be provided to protect the privacy of others using the AI.

In terms of transparency, users should be able to see how their data are being used to compute specific work outcomes. For example, the AI should be transparent to Sue and other workers concerning data on compensation that are being used and how additional data such as maternity/paternity leave were computed into compensation evaluations. Additionally, the AI system should be transparent in terms of what data from comparable groups are being included in the computational model when determining outcomes such as Sue's compensation. Transparency also affords Sue the ability to act upon this information if she finds that relative to her comparable peers, she has been paid unfairly. Specifically, if Sue wanted to address that she had been paid unfairly, the AI should be transparent so that Sue can see how much she would have earned if she had been paid similarly to her peers with commensurable skills and tenure at the company. Nonetheless, designers should consider the tradeoffs with privacy when considering the transparency affordance. For instance, the organization might not want to violate other workers' privacy by sharing information about their earnings or whether they accepted or declined a leave.

Additionally, the explainability affordance would be important in the AI describing to Sue how it arrived at the decision for her compensation changes based on her sales performance data. Sue would then have an explanation that was communicated to her in understandable terms that related to her compensation outcomes. In this way, if Sue believed her compensation decision was unfair, she would at least have the understanding to contest an unfair outcome. Finally, to provide Sue with additional insights about her compensation outcomes, the visualization affordance would be important concerning how aggregated and raw data are presented for the purposes of benchmarking her compensation outcomes relative to others'.

**AI fairness example case: Procedural fairness**

Procedural fairness is often discussed in instances when organizations use established protocols and policies to coordinate employees, such as with scheduling work shifts for hourly workers in service industries. The number of hours an employee works directly impacts his or her earnings. Also, whether employees are scheduled during peak sales hours can affect their earnings in tips and commissions. For instance, consider Brandon, a salesperson for a retail store whose earnings are based on the number of hours worked and commissions on each sale he completes. Thus, Brandon prefers to be scheduled during peak hours when consumer foot traffic is high in the store. Brandon's manager, Charles, plans the weekly schedule for the store based on employees' availability (i.e., accuracy). Although the company policy states that managers should not make scheduling decisions based on favoritism, Charles sometimes disproportionately assigns better work shifts to Dennis (i.e., unbiased suppression) because Charles is more aware of Dennis's scheduling constraints. Thus, Charles adjusts the employee schedule to account for the factors that would prevent Dennis from coming to work (i.e., correctability). Recognizing that employees like Brandon might be unfairly scheduled for less desirable work shifts, the company could correct for procedural unfairness by ensuring that the company policy is applied consistently and accurately for all employees, not just for those who are personally favored.

AI is one potential solution the company might use to automate when employees are scheduled for their work shifts (Mateescu & Nguyen, 2019). The AI could account for all employees' availability rather than prioritizing only a few favored employees based on their personal relationships with the manager. However, the AI could still threaten procedural fairness if the system were unable to account for informal procedures that are still important when scheduling employees. For instance, there might be informal procedures managers do such as checking employees' personal constraints—such as lacking reliable transportation or having to manage child care responsibilities—when assessing employees' availability for the work schedule. As such, when designing AI for procedural fairness, designers also need to consider how the formal and informal procedures can shape procedural fairness outcomes.

*Operationalizing fairness*

The first step in operationalizing procedural fairness involves deciding whether the current procedures are fair to employees in the first place according to the previously discussed characteristics of procedural fairness (consistency, unbiased suppression, representativeness, correctability, accuracy, ethicality). Embedding an unfair procedure into an AI system only exacerbates unfair outcomes for employees. Next, even if a procedure is determined to be fair, it might not be amenable to an AI system if implicit knowledge is required to enact the procedure. As alluded to in Brandon's case, there might be informal procedures such as considering employees' personal constraints when determining their availability to work certain shifts. To this point, Lee (2018) also showed that people thought that AI could be as fair as human managers for decisions based on explicit knowledge but that AI could not be fair when decisions were based on tacit knowledge that only a human manager might possess. Thus, procedures that also require tacit knowledge might require using both human and AI in tandem to consistently produce procedurally fair outcomes. For the purposes outlined in Brandon's case, we could (1) operationalize that the existing organizational procedure scheduling workers by their actual availability and not by personal preference for peak hours is a fair procedure and (2) operationalize that the scheduling procedure delegated to the AI should only account for the explicit knowledge provided by employees concerning their hourly availability.

*Data and model selection*

This operationalization of procedural fairness raises important questions related to the data and models selected for the AI, including: (1) Who decides what is deemed as correct, accurate, unbiased data? and (2) What computational model consistently and ethically represents the specific steps of the given procedure in the AI? In the case of Brandon's work availability, the AI would still require Brandon and other employees to input their available work hours. Even though the AI, not the human manager, would schedule the work shifts, there might be reason to suspect employees would still work around the system by only telling the system they were available during peak hours in order to compete for the more desirable work shifts. However, if someone else were to input an employees' availability on their behalf, this might not represent correct or accurate data.

With regard to selecting the computational model, designers would need to determine how to consistently and ethically designate the specific steps of the procedure in the AI. For instance, when Brandon and Dennis both indicate that they are available for the same work hours, the AI could randomly assign them so there is no preference in who is scheduled during peak hours; alternatively, the AI could prioritize who gets which work shift based on prior sales performance data so that high performers are scheduled during peak hours to generate higher revenue.

*Evaluating procedural fairness*

To evaluate procedural fairness, designers should consider whether: (1) all steps in the procedure have been completed consistently, correctly, and accurately; (2) all parties have been heard before a decision has been made; and (3) the system allows for the procedure to be corrected. For instance, in the scheduling case described here, if Brandon or Dennis did not input their availability for the work week, the AI could automatically notify them that this step was missed. Similarly, the AI could indicate whether there were any instances of inaccurate data (e.g., inputting available work hours when the store is closed) and notify the manager if some employees' availability data are missing. Because representativeness and correctability are also characteristics of procedural fairness, there might still need to be manual, human intervention allowed in the AI. For example, to evaluate whether the AI-produced schedule is indeed fair, the AI scheduling system might only be finalized and distributed to employees after Charles's approval as a manager.

*Redressing instances of procedural unfairness*

Redressing becomes important when the procedure has not been followed or that procedure has been shown to be unfair in practice. This raises questions concerning how AI can be designed:

(1) to consistently follow existing procedures while being flexible enough to correct for unfairness in the set procedure and (2) to provide workers with alternative procedures to achieve fairness. In cases where the steps in the procedure were incorrectly or inconsistently followed, the worker could be given another opportunity to access the procedure. For example, if Brandon was not scheduled for any peak hours for the upcoming work schedule, the AI might prioritize him for peak hours in the following week's schedule. In cases where the procedure has been shown to be unfair, redressing becomes more complex because it involves having the AI follow another procedure. For instance, Brandon and other employees might still be scheduled during fewer peak hours if the AI scheduling system assigns peak hours randomly because it would be contingent on other employees' more restricted availability. Thus, management might decide to follow another procedure, such as prioritizing who is scheduled for peak hours based on performance or seniority in addition to availability. Nonetheless, redressing for procedural unfairness could also require removing the AI system from the given organizational procedure altogether if the AI continued to produce unfair or biased outcomes.

*Affordances to support procedural fairness*

Transparency, explainability, and voice play an important role supporting procedural fairness in AI systems. The fundamental tenets of procedural fairness require having a transparent process, ensuring employees actually understand, and providing opportunities for all employees to be heard in matters related to the given procedure. As such, this raises important questions for designers, including: (1) What metrics should be used to determine whether the procedure is transparent in the AI?, (2) How can AI systems increase employees' awareness of the given procedure?, (3) How can AI provide feedback to employees concerning the past and future actions in the procedure?, and (4) How can AI offer employees a voice in how procedures are carried out and help employees raise concerns when they experience procedural unfairness?

At a basic level, AI should ensure that all details concerning the process are transparent to employees. For instance, in our problem case with Brandon, the AI should be transparent about how the system decides when to schedule employees' work shifts. The AI should be transparent about how Brandon and Dennis's hours were scheduled and when Charles approves or changes the final work schedule. However, workers might acknowledge that they are aware of the procedure but do not understand how the procedure works in practice until they are reprimanded for not abiding by the given procedure. In this regard, the AI could measure the extent to which employees understand the procedure by requesting self-reports or using other relevant metrics (e.g., weekly errors made following the scheduling procedure) to assess understanding of the procedure. Furthermore, designers should ensure that the potential consequences for not following procedures are also transparent in the AI system. Consider the scenario where Charles requests changes to the employee work schedule to favor Dennis over Brandon. The AI could support transparency by using feedforward systems that make it clear what the desirable behaviors are (e.g., ensuring all employees have an equal chance to work during peak hours) rather than only enacting negative consequences when employees do not follow the procedure. Last, in terms of transparency, the AI should be designed to provide feedback to inform workers what actions in the procedure have already been completed and to anticipate future actions that are still needed to complete the procedure. For example, if Brandon were to submit a grievance because he believes the procedure is not fairly scheduling his work hours, the AI could inform him that management has acknowledged that his grievance has been submitted and how it has decided to redress his grievance.

In addition to transparency, the AI system should include explainability as an affordance to inform employees when and why their work schedules have been changed according to the company's scheduling procedures. For instance, rather than only notifying Brandon that his work schedule has changed, the system could also explain to Brandon why his work schedule has changed based on the availability information he provided to the system and how time schedules are determined by peak hours for customer traffic.

Finally, the affordance of voice would enable Brandon the opportunity to contribute information indicating his weekly availability to work. After his manager signs off on the final work schedule, the voice affordance would also provide Brandon the chance to give feedback to the AI indicating when he believes he was unfairly scheduled to miss peak customer hours. Furthermore, the voice affordance would provide Brandon the opportunity to agree with or raise his concerns about the scheduling procedure itself. However, as previously discussed on measuring the effectiveness of affordances, Brandon would need to feel empowered when providing information about his availability and sending feedback when he is overscheduled during non-peak hours to believe that his contributions would result in fair scheduling outcomes.

**AI fairness example case: Interactional fairness**

Interactional fairness focuses on how organizations respect and inform their workers. Consider a case in which two employees, Adam and Beth, are participating in their company's annual performance review with their manager, Cathy. Performance reviews can be time-consuming and labor-intensive. The process requires Cathy to assess Adam and Beth's skills and potential for a promotion—along with other employees in their department—as objectively as possible. Although Cathy has several years of experience completing performance reviews, it is still challenging for her to do them objectively. Cathy previously worked in Adam's position before becoming a manager and is more lenient in assessing Adam than Beth. Moreover, there might be aspects of Adam and Beth's work that are invisible to Cathy, such as the work Beth completes after hours at home or how Adam contributes to a positive work culture. To help deal with the information overload, Cathy uses the metrics and rating systems on the performance review to systematically codify Adam and Beth's performance. Also, Cathy meets with Adam and Beth in person to inform them about their performance review outcomes.

If the company were to use an AI, this could be an opportunity for Cathy to offload the information overload and reduce human biases in conducting performance reviews. Additionally, the AI could assess Adam and Beth over time and consider other important data points on their skills and work outcomes that enable the AI to predict whether Adam and Beth are candidates for promotion or retention, or are at risk for turnover. However, the AI might present challenges in conducting performance reviews with interactional fairness. For one, AI can collect enormous volumes of information about workers and presenting all of that available information can overload or distract workers from relevant information about their performance review. There is also the risk that the AI lacks informational fairness if it does not inform Adam and Beth with adequate explanations about their performance in the same manner as Cathy, who has a shared human experience working at the company. Furthermore, Adam and Beth might perceive the AI as impersonal when they receive their performance review evaluations and be unable to communicate about their career paths in the ways they would with Cathy (i.e., interpersonal fairness).

*Operationalizing fairness*

Both components of interactional fairness—informational and interpersonal—need to be reflected in our operationalization of interactional fairness in order to design AI applications as non-human agents that are considerate, respectful, and truthful to human users. First, with informational fairness, Reeves and Nass's (1996) characterization of "polite" technology provides a lens to operationalize informational fairness in that technology should only give information that is necessary, truthful, and relevant. Second, with interpersonal fairness, Cooper, Reimann, and Cronin (2007) suggested that characteristics of considerate interactive technological systems provide a lens to operationalize interpersonal fairness in terms of accommodating the goals and needs of users. A considerate system goes beyond performing basic functions, making the needs of its users a key concern. For the purposes outlined in the problem case with Adam, Beth, and Cathy, we can operationalize interactional fairness in terms of (1) whether workers are provided clear information about the AI (Abrams et al., 2019; Bobocel & Farrell, 1996; Shapiro et al., 1994) and (2) whether the AI treats workers with respect and encourages cooperation in the workplace (Cropanzano et al., 2007).

*Data and model selection*

One of the focuses of interactional fairness centers on treating employees with respect. Thus, when considering data and model selection for an AI application, designers should address questions concerning (1) how much data employees can access through the AI and (2) what data are needed in order to be considerate across different groups and not just the majority groups. In their framework for ethical AI systems, Abrams et al. (2019) highlighted the importance of complete and accurate information in interactions between AI systems and the people who use them. The first step in designing AI features that present adequate information to the worker is to determine what "complete and accurate" means for the workers. However, it is important to consider what information workers might want versus what they actually need. For example, in our problem case this might entail that the AI system provide Adam and Beth access to all of their performance data over time so they can develop a shared understanding with their manager Cathy about their career progress. Adam might also want to know Beth's performance review evaluations to understand where he ranks in relation to his colleagues, but that comparative information might not serve him well in improving his individual performance and providing such information would also violate Beth's privacy.

Additionally, designing AI systems to support interactional fairness requires the technology to be considerate to all employees. As such, the data selected for the AI system should not be biased toward the majority or a particular group so that the AI system treats all employees fairly. For instance, in our problem case, explanation associated with the performance review AI system might need to differ in tone, language, and information depending on the user. If the AI system was trained on data that favored employees more similar to Adam, it might not interact in ways that are perceived as considerate to employees who are more similar to Beth. This is of particular concern in organizational contexts where employees are working across different cultures or language preferences.

*Evaluating interactional fairness*

To evaluate interactional fairness, designers should consider whether (1) workers are satisfied and understand the information presented by the AI system and (2) workers feel respected by the AI as a non-human agent. For example, the performance review AI system might present far more detailed information than Cathy was capable of providing about the review process and how the employee's performance was analyzed. On one hand, Adam and Beth might be satisfied with the detail and amount of information provided. On the other hand, if the information is presented in a way that makes it difficult for Adam or Beth to understand the meaning of that information or they lack ways to receive clarification about the information presented, then the AI system might not be deemed fair in terms of interactional fairness. Moreover, the AI system would need to be evaluated in terms of how respectful it is as a non-human agent. For instance, if Adam received a negative performance review, he might feel disrespected in receiving poor news from a non-human agent. Although designers of the AI performance review system might have attempted to design a non-human agent that is considerate and empathetic, the interpersonal component of interactional fairness can only be evaluated through the perspective of the workers using the system. As such, designers need to evaluate interactional fairness by allowing users to provide feedback on their experiences interacting with the system.

*Redressing instances of interactional unfairness*

Through the lens of interactional fairness, AI systems should focus on redressing issues related to informational and interpersonal unfairness. This in turn raises important questions concerning whether (1) explanations provided by the AI system are effective at redressing issues of informational unfairness and (2) socially sensitive AI systems (i.e., systems that display emotions or regret) are effective at redressing issues of interpersonal unfairness.

For instance, the AI system might have hidden or made it difficult to find information related to all the dimensions in which employees are evaluated for their performance reviews. With regard to providing effective explanations to redress information unfairness, the AI system could directly redress this instance of unfairness by providing workers immediate access to previously unavailable information such as the metrics used for performance evaluations. If there is information that cannot be provided to workers, such as anonymous feedback from colleagues, the system could redress this information unfairness by providing an explanation to clarify that it is not accessible to workers.

For interpersonal unfairness, redressing is particularly important for maintaining positive relationships between two parties (Goodstein & Aquino, 2010; Skarlicki & Folger, 1997). HCI researchers have studied ways to ensure that technology is polite and considerate in interacting with humans (Cooper et al., 2007; Reeves & Nass, 1996), such as designing systems that are deferential and take responsibility for social violations. One approach to redressing interpersonal unfairness is to have the AI apologize if users perceive the technology to be rude or inconsiderate (Kidder, 2007; Skarlicki & Latham, 1996). The effectiveness of such redressing actions might be improved if the AI were designed to show social sensitivity to the worker (Skarlicki & Folger, 1997). For example, this might require the AI to use specific words or change its tone when delivering negative news about a poor performance review to Adam. Nonetheless, it is also important to consider that some users might find it socially insensitive to use an AI system for delivering personal information such as the outcome of one's performance review.

*Affordances to support interactional fairness*

For interactional fairness, the primary affordances that need to be considered are transparency, explainability, and visualization. First, the transparency affordance could provide Adam and Beth access to see the data that are being used to assess their performance. Second, the explainability affordance could provide Adam and Beth detailed summaries show how their data were used in their review assessment. These explanations could also be useful for Adam and Beth in making informed decisions about how to improve their performance based on the ways in which the performance review system evaluates their work. Also, if Adam or Beth wanted to discuss their performance review with their manager Cathy, they would have an understanding of how they were evaluated. The visualization affordance addresses the informational component that concerns balancing the amount of information presented to users: too much information might not allow workers to understand and act upon that information. Last, the voice affordance can support the interpersonal component of interactional fairness—Adam and Beth should have a way to communicate back to the AI if they believe their performance has been unfairly evaluated. Therefore, designers of AI systems should consider these questions: (1) How much autonomy should workers have in adjusting the level of detail and curating information for their specific needs?, (2) How often should information be presented to users?, (3) How should information be shown objectively to workers?, and (4) How should AI systems promote feelings of respect?

For instance, Adam and Beth might have different areas of their performance reviews that they want to track over time, thus the AI system should afford them the autonomy to visualize their data in ways that make sense to their informational needs. Similarly, designers should consider the tradeoffs of affording information transparency while not overburdening users with information updates that are distracting or introduce information overload at work. Furthermore, the information presented on the AI system should not be misrepresented. For instance, if Adam needs clarification about why he received a low performance review, the AI system should present information to him objectively and not unduly influence him to assume that more transparent data means his performance review was evaluated fairly. Last, affording workers a voice in the AI system is one way in which the technology can promote feelings of respect through redressing instances of interactional unfairness. For example, the AI system could afford Adam a way to express his grievance about the specific ways in which he was evaluated on his performance review. The previous affordances of transparency and visualization should also support the ways in which Adam can express his voice through the system by being able to make a case based on the information available to him.

# 5. Future Work and Limitations

Our design agenda on artificial intelligence (AI) fairness was intended to identify crucial design issues through the lens of our theoretical framework. Nonetheless, several other design issues are not covered in our agenda. Each of these issues would justify the need for its own design agenda relative to AI fairness. These issues are privacy, autonomy, organizational context, wider perspectives of fairness, equity vs. equality, AI accountability, and AI audits and auditability. To avoid inadequate coverage, we present and discuss each issue as a limitation of our design agenda, and opportunities for future directions (see Table 6).

**Table 6**
**Future Work and Limitations**

| Design Issue | Primary Question |
|---|---|
| Protecting Worker Privacy | How can employee privacy be protected? |
| AI Autonomy | How much autonomy should the AI be given? |
| Organizational Context | How can the organization's policies, priorities, and culture be included in the AI's design? |
| Fair to Whom? | How can fairness from the perspectives of the organization, employee, and the customer all be integrated? |
| Equity vs. Equality | Should AI fairness be determined by equity or equality? |
| AI Accountability | Who should be held legally and financially accountable for the AI's actions? |
| AI Audits and Auditability | How can an AI be designed to allow or support the ability to be audited? |

*Note.* AI = artificial intelligence

## 5.1. Protecting Worker Privacy

The first of these issues is worker privacy. Privacy is often defined by whether people can exclude themselves from being observed. Many aspects of our design agenda require access to user data. In many cases, to assess fairness, data are needed from more than one user. For example, to assess distributive fairness, comparison data from similar other workers are needed. This requires permission to obtain data from many others to assess fairness for the one. Another important element of privacy concerns the need for anonymity for those seeking fairness. For example, an individual who files a complaint of unfairness needs some layer of protection to avoid being personally identified and targeted. Ultimately, future design agendas that focus entirely on the need to balance data requirements with privacy should be developed.

## 5.2.   AI Autonomy

Another important design question concerns the degree of autonomy given to the AI. How much autonomy should the AI have? For example, should the AI only be allowed to identify instances of unfairness? When should a human manager be contacted? Should the AI identify and recommend a course of action? Should the AI wait for permission from a human manager to approve any action? Or should the AI automatically identify, select, and immediately take a course of action? We can also imagine design recommendations where the AI would be given more or less autonomy depending on the situation or fairness type. A research agenda could focus on determining what elements of a given situation should be used to determine the degree of AI autonomy. Another question not to be overlooked is whether the AI should ever be involved in issues related to fairness. This is much more of a philosophical question because our design agenda clearly assumes that AI should be involved. Nonetheless, a future design agenda could seek to understand the inherent appropriateness and limitations of AI in organizational fairness.

## 5.3. Organizational Context

Our design agenda employs a strong theoretical framework of fairness that leverages similarities among organizations. Nonetheless, future design agendas should consider the differences among organizations. For example, an organization's policies, priorities, and culture are likely to impact design in at least two ways. One, they are likely to influence which type or definition of fairness an organization chooses to promote. Two, an organization's policies, priorities, and culture are likely to enable and constrain the effectiveness of any designs related to AI fairness. Recently, organizational scholars have begun to pay more attention to the issues of AI as they pertain to organizational policies (see Daugherty et al., 2019). Therefore, future design agendas that systematically consider organizational context could start by drawing from the work of organizational scholars.

## 5.4. Fair to Whom?

Our design agenda takes the perspective of fairness from the individual being managed by the AI. However, a design agenda could be developed from the perspective of a customer. We can imagine scenarios where the AI is interacting with customers. From this view, distributive fairness could be defined, for example, by whether the AI ensures that women do not pay more for a service or product than men. Price differences between men and women are not uncommon (Ayres & Siegelman, 1995). Another view is that of the organization itself. AI systems could be designed to ensure that neither workers nor customers are unfair to the organization. Yet another perspective that is particularly relevant for platform companies is the tension between fairness for the service provider and the customer. For example, an Uber AI system must balance distributive fairness for both the driver and the rider. In all, new design agendas could be derived entirely by changing the perspective of the entity one seeks to promote fairness for in a situation.

## 5.5. Equity vs. Equality

There has long been a debate on the merits of equity versus equality. To be clear, the definition of *equity* in reference to Adams' equity theory does not translate well to this discussion. In the debate on equity versus equality, equity focuses on providing individuals what they need to be successful. This often requires treating individuals differently based on their needs. Equality focuses on treating everyone the same regardless of their individual needs. Equality assumes that everyone starts with the same needs. Once again, despite the word *equity* in Adams' equity theory, the theory seeks to promote fairness by equality and so does organizational justice theory. Hence, our design agenda is also based on fairness through equality rather than equity. To the degree that individuals have similar needs, equality is probably the preferred approach. However, if individuals have widely different needs, equity might be the preferred approach.

## 5.6. AI Accountability

AI accountability is an emerging and important area in the study of the implications of AI within and outside organizations (Wachter, Mittelstadt, & Floridi, 2017). For clarity, we define AI accountability as the assigning or identification of particular persons, groups, or organizations that can be held legally and financially accountable for the AI's actions. In this paper, we assumed that the organization that employs the worker holds primary accountability for its actions. However, we acknowledge that this is not always true. For example, who should be held accountable when an organization outsources its hiring to a third-party firm that deploys a biased or unfair AI application? This problem is made harder when the variables selected in the model were chosen by the third-party hiring firm's AI but the data were provided by the organization. Such a situation would make it difficult to determine who should be held accountable because the bias could have resulted from the variables selected in the model, the sample employed, or a combination of both. Another just as problematic situation is when outcomes are the result of decisions of several AIs deployed by several organizations. Even when the source of the problem can be detected, issues of legal and financial accountability are not always directly tied to the source of the problem. To that end, future research is needed to specifically address issues of AI accountability that include intimate knowledge of the organization's legal environment.

## 5.7.    AI Audits and Auditability

One approach to assess the degree of potential bias in AI is to conduct an audit. There are many definitions of AI audits (LaBrie & Steinke, 2019; Sandvig, Hamilton, Karahalios, & Langbort, 2014). In this paper, an AI audit is defined as an inspection of the AI's underlying logic, decision criteria, and data sources in an attempt to validate the AI. AI validation is typically conducted to demonstrate that the AI complies with current laws, regulations, or policies. AI auditability, as an affordance, is the extent to which the AI allows or supports the ability to be audited. AI audits can be conducted manually via logical walk-throughs, automatically via code checks, or by large-scale computer simulations. AI audits can also be conducted periodically to ensure that the AI is still in compliance.

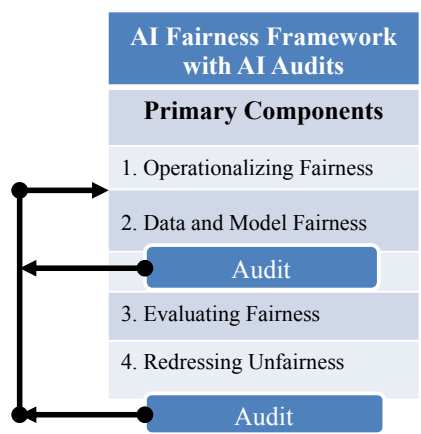

**Figure 2. Artificial intelligence (AI) fairness framework with AI audits**

As shown in Figure 2, Steps 1 and 2 provide the groundwork for any potential audits. That being the case there are also two places where AI audits could be used more explicitly in our proposed framework. First, after ensuring data and model fairness in Step 2, a simulation could be conducted to assess the fairness of the potential outcomes before deploying the AI. This would essentially allow the organization to evaluate the outcomes associated with the AI without actually deploying it. Second, after redressing unfairness, Step 4, another audit could be conducted to track what occurred that led to the AI unfairness (see Figure 2).

Our current framework does not explicitly include AI audits as a formal step. Unlike the affordances mentioned in this paper, AI audits and auditability, as an affordance, is not directed toward employees but primarily directed at the organizational leadership in an attempt to ensure that the organization is meeting specific external requirements. However, those specific requirements are highly dependent on the organization's legal and regulatory environment. The proposed framework in this paper largely assumes that organizations have the primary say on what is or is not fair (i.e., Step 1). The degree to which AI audits are directed to meet legal and regulatory requirements where others outside the organization determine what is and is not fair the framework would need to be supplemented. However, the degree to which AI audits are performed for internal purposes can be used to incorporate AI audits into our proposed framework.

## 6. Conclusion

In this paper we pushed forward the design of artificial intelligence (AI) fairness in organizations by reviewing existing literature, employing a strong theoretical framework, and proposing a design agenda. On the basis of the literature review, which included 25 design papers, we showed that the literature has often overlooked differences among types of fairness. The literature has instead focused more on what we have identified as algorithm fairness. Further, after classifying the literature based on the organizational justice theoretical framework, we found that more attention has been paid to distributive fairness and less to interactional fairness. In all, we hope our design agenda helps to overcome these limitations and provides a starting point for design.

## Appendix. Search Process for AI Fairness Literature

**Study Selection Process**

We employed several search engines for the literature review: Google Scholar, ACM Digital Library, AIS Digital Library, Scopus, PsycINFO, and IEEE Xplore. We conducted the search between April 2019 and May 2019 using three main groups of search terms. The first group included the words "artificial intelligence, AI, AI management systems"; the second included "fair, fairness, justice, organization justice"; and the third group of search terms included "organization, work, employment, worker." These search terms were used across all search engines.

The search terms returned thousands of potential articles presented in order of relevance to the topic. We refined the search by going through each article based on the initial inclusion criteria until the following page of listings yielded no relevant articles. The articles returned beyond this point only included terms related to one of the search terms—AI or fairness or organizations. The search based on the initial inclusion criteria yielded 61 unique articles.

**Initial inclusion criteria**. We initially included papers if they explicitly or implicitly discussed at least one of the three fairness types. We only included papers that were published in English-language journals or conference proceedings. We then screened the 61 articles against the final inclusion and exclusion criteria described here. This resulted in 25 publications, of which 16 were empirical papers and nine were conceptual papers; there were eight journal articles and 17 conference proceedings published between 2008 and 2019. See Table 2 for a summary of the papers by time period and Table 3 for a summary of the papers by fairness type.

**Final inclusion criteria**. We included papers if they satisfied two criteria: (1) they discussed issues related to AI fairness in organizations or workplaces and (2) they explicitly discussed design implications. Papers that discussed one of the fairness types without explicitly mentioning a specific fairness type term were still included in final set of 25 publications.

**Exclusion criteria**. We excluded papers from the final set of publications if they met any of these three criteria: (1) they did not mention issues related to fairness in AI systems, (2) they did not apply to an organization or workplace, or (3) they did not discuss any design implications.

## Coding Process

To understand how the AI fairness literature incorporates organizational justice theory when designing AI systems, we coded the articles from our literature search into the fairness type (distributive, procedural, interactional) and justice type (retributive and restorative). Using the definitions of fairness and justice types previously summarized, three researchers independently reviewed every paper and assigned them a fairness and justice type. Papers could be assigned multiple fairness and justice types. The research team then discussed each paper and resolved any disagreements in the coding process until we reached consensus. Table 3 summarizes how we organized the literature by fairness types.